\documentclass[11pt]{article}

\def\a{\alpha}
\def\b{\beta}

\def\d{\delta}
\def\e{\epsilon}
\def\f{\varphi}
\def\g{\gamma}
\def\h{\eta}

\def\j{\psi}
\def\k{\kappa}
\def\l{\lambda}

\def\q{\vartheta}

\def\s{\sigma}
\def\t{\tau}

\def\x{\xi}

\def\D{\Delta}
\def\F{\Phi}

\DeclareFontFamily{OT1}{msb}{}{}
\DeclareFontShape{OT1}{msb}{m}{n}
 {  <5> <6> <7> <8> <9> <10> gen * msbm
      <10.95><12><14.4><17.28><20.74><24.88>msbm10}{}
\DeclareMathAlphabet{\bubble}{OT1}{msb}{m}{n}

\newfont{\goth}{eufm10 scaled \magstep1}
\def\cg{\mbox{\goth g}}

\def\ch{{\cal H}}

\def\cl{{\cal L}}

\def\cp{{\cal P}}

\def\cu{{\cal U}}
\def\cv{{\cal V}}

\def\fr#1#2{{\textstyle{#1\over #2}}}      
\def\p{\partial}

\def\dtt#1{{\buildrel {\hbox{\,\LARGE .}} \over {#1}}}
\def\dttt#1{{\buildrel {\hbox{\hskip3pt\LARGE .}} \over {#1}}}
\def\dt#1{{\buildrel {\hbox{\hskip-2pt\LARGE .}} \over {#1}}}
\def\ddt#1{{\buildrel {\hbox{\,\LARGE .\hskip-2pt.}} \over {#1}}}
\def\phip{{\buildrel {\hbox{}} \over {\phi}}'}

\def\square{\kern1pt\vbox
            {\hrule height 0.6pt\hbox{\vrule width 0.6pt\hskip 3pt
 \vbox{\vskip 6pt}\hskip 3pt\vrule width 0.6pt}\hrule height 0.6pt}\kern1pt}

\def\ra{\rightarrow}

\def\der#1{{\textstyle{\partial \over \partial #1}}}

\def\be{\begin{equation}}
\def\ee{\end{equation}}
\def\la#1{\label{#1}} 
\def\re#1{(\ref{#1})} 
\def\arr{\begin{array}{rlll}}
\def\ea{\end{array}}
\def\bea{\begin{eqnarray}}
\def\eea{\end{eqnarray}}  

\catcode`@=11
\@addtoreset{equation}{section}
\catcode`@=12

\begin{document}
\begin{titlepage}
\rightline{solv-int/9811016}
\vskip 1.5 true cm
\begin{center}
{\Large  The supersymmetric Camassa-Holm equation \\[5pt]            
and geodesic flow on the superconformal group}
\vskip 1.5 true cm
{\large Chandrashekar Devchand$\;^{1,2}$\ \ and\ \ Jeremy Schiff$\;^3$}
\vskip 0.8 true cm
{\it $^1$ Max-Planck-Institut f\"ur Mathematik}\\
{\it Gottfried-Claren-Stra\ss{}e 26, 53225 Bonn, Germany}\\[5pt]
{\it $^2$ Max-Planck-Institut f\"ur Mathematik in den Naturwissenschaften}\\
{\it Inselstra\ss{}e 22-26, 04103 Leipzig, Germany}\\
{\small devchand@mis.mpg.de}\\[5pt]
{\it $^3$ Department of Mathematics and Computer Science }\\
{\it Bar--Ilan University, Ramat Gan 52900, Israel}\\
{\small schiff@math.biu.ac.il}
\end{center}
\vskip 1.5 true cm
\begin{quote} 
\centerline{\bf Abstract}
We study a family of fermionic extensions of the Camassa-Holm equation.
Within this family we identify three interesting classes:
(a) equations, which are inherently hamiltonian, describing  geodesic flow 
with respect to an $H^1$ metric on the group of superconformal transformations
in two dimensions, (b) equations which are hamiltonian with
respect to a different hamiltonian structure and (c) supersymmetric
flow equations. Classes (a) and (b) have no intersection, but the 
intersection of classes (a) and (c) gives a candidate for a new
supersymmetric integrable system. We demonstrate the Painlev\'e property 
for some simple but nontrivial reductions of this system.
\end{quote}
\vfill
\end{titlepage}

\section{Introduction}
Recently there has been substantial interest in the  Camassa-Holm (CH) 
equation \cite{ch1,ch2}:
\be
u_t - \nu u_{xxt} = \k u_x - 3uu_x + \nu (u u_{xxx} + 2 u_x u_{xx}) \ .
\la{ch}\ee
This equation has been proposed as a model for shallow water waves. It
is believed to be integrable (having bihamiltonian structure), but due to 
the nonlinear dispersion term, $uu_{xxx}$, it exhibits more general wave 
phenomena than other integrable water wave equations such as KdV. In 
particular it admits a class of nonanalytic weak solutions known as 
{\em peakons}, as well as finite time blow-up of solutions.

Geometrically, the relationship of CH to KdV is rather deeper: 
Both are regularisations of the Euler equation for a
one dimensional compressible fluid (Monge or inviscid  Burgers equation), 
\be
u_t = - 3uu_x\ .
\la{monge}\ee
This latter equation describes geodesic motion on the group of 
diffeomorphisms of the circle  Diff($S^1$) \cite{arn} with respect to a 
metric induced by an $L^2$ norm, $\ \int u^2 dx\ $, on the associated 
algebra. If the group is centrally 
extended to the Bott-Virasoro group, the KdV equation arises
\cite{ok,mis2,arn3,michor}. On the other hand, if the metric is changed to 
one induced by an $H^1$ norm, $\ \int (u^2 + \nu u_x^2) dx\ $, 
the CH equation arises \cite{mis,sh,ko}. Both these `deformations'
have a regularising effect on solutions of \re{monge}, which exhibit
discontinuous shocks. 

Thus KdV and CH arise in a unified geometric setting; both are 
integrable systems which describe geodesic flows.
This raises an important question: What features of the underlying
geometry give rise to integrability? 
In general, geodesic flows of this type are {\em not} integrable: the 
Euler equation for fluid flow in more than one spatial dimension is 
an example \cite{arn}. Indeed, for the latter, Arnold has suggested 
a relationship between negative sectional curvatures 
and non-predictability of the flow. Is integrability also 
geometrically determined? 

One further example of an integrable bihamiltonian system arising from
a geodesic flow has been discussed in the pioneering paper of Ovsienko 
and Khesin \cite{ok}. Using the superconformal group with an $L^2$ type
metric, they obtained the so-called kuperKdV system of Kupershmidt
\cite{k}. This is a fermionic extension of KdV: it describes evolution 
of functions valued in (the odd or even parts of) a grassmann algebra. 
In fact, as we will see below, taking a general $L^2$ type metric 
on the superconformal group gives rise to a one parameter family of fermionic
extensions of KdV, which includes not only kuperKdV, but also the superKdV
system of Mathieu and Manin-Radul \cite{m,mr}. The latter is integrable:
it has only a single hamiltonian structure, but unlike kuperKdV it is
supersymmetric, a property which is widely believed to contribute to 
integrability. It remains a mystery as to why, of the one parameter
family of geodesic flow equations associated with $L^2$ type metrics on the 
superconformal group, only two specific choices of the parameter
give rise to integrable systems.

The main purpose of this paper is to investigate geodesic flow equations
obtained from $H^1$ type norms on the superconformal group; more generally
we consider the following family of fermionic extensions of CH:
\bea
u_t - \nu u_{xxt} &=&  \k_1 u_x +\k_2 u_{xxx} + \b_1 uu_x 
                   + \b_2 u_xu_{xx} + \b_3 uu_{xxx}
    + \g_1 \x\x_{xx} + \g_2 \x_x\x_{xxx}+ \g_3 \x\x_{xxxx}
\nonumber\\[6pt]
\x_t - \mu \x_{xxt} &=&  \s_1 \x_x + \s_2 \x_{xxx} +\e_1 u_x \x + \e_2 u \x_x
            + \rho_1 u\x_{xxx} + \rho_2 u_x \x_{xx} + \rho_3 u_{xx}\x_x
            + \rho_4 u_{xxx} \x  \ .
\la{gen}\eea
Here $u(x,t)$ and $\x(x,t)$ are fields valued, respectively, in the even 
and odd parts of a grassmann algebra, and
$\ \{\nu,\mu,\k_1,\k_2,\b_1,\b_2,\b_3,\g_1,
\g_2,\g_3,\s_1,\s_2,\e_1,\e_2,\rho_1,\rho_2,\rho_3,\rho_4\}\ $ are 
parameters. By rescaling $u$ and $\x$ it is possible to set $\,\b_1{=}-3\,$ 
and $\,\g_1{=}2\,$ (assuming that they are nonzero), and we shall do this 
throughout. In addition it is possible to eliminate up to two further
parameters by rescaling  the coordinates $x,t$.

We derive three interesting classes of systems of the form \re{gen}. 
In section 2, we consider geodesic flow on the superconformal group
with an $H^1$ type metric; the resulting systems have a natural hamiltonian
structure, or more precisely, since the fields are grassmann algebra valued,
a graded hamiltonian structure. In section 3 we identify a class of systems 
having a different hamiltonian structure. Unfortunately the latter has no 
intersection with the class of section 2, so there does not seem to be a 
bihamiltonian fermionic extension of CH. 
In section 4 we consider systems of the form \re{gen} that are
invariant under supersymmetry transformations between $u$ and $\x$.
This class has nontrivial intersections with both the classes of sections 
2 and 3. In particular, there is a unique  supersymmetric 
geodesic flow system, which is a promising
candidate for being integrable. In section 5 we show that two
reductions of this system have the Painlev\'e property.

A trivial integrable CH system of the form \re{gen}, which is not incorporated
in the classes of sections 2,3, and 4, and which we shall not discuss further, 
is the odd linearisation of the bosonic CH system \re{ch}
\be\arr
u_t - \nu u_{xxt} &=& \k u_x  - 3uu_x + \nu  (u u_{xxx} + 2 u_x u_{xx})\ ,\\[6pt]
\x_t - \nu \x_{xxt} &=& \k \x_x  - 3(\x u )_x 
                     +   \nu (\x u_{xxx} + u \x_{xxx}+ 2 (\x_x u_x )_x)\ .
\ea\la{lin_ch}\ee
Replacing $\,u\,$ by $\,u+{\k\over 3}\,$ and considering the limit 
$\,\nu\rightarrow 0\,$, $\k\rightarrow\infty$, with $\nu\k=3$, yields the system
\be\arr
u_t  &=& - 3uu_x + u_{xxx}\ ,\\[6pt]
\x_t &=& - 3(\x u)_x + \x_{xxx}\ .
\ea \ee
This trivial fermionic extension of KdV has appeared often in 
the literature (see e.g. \cite{m}). 

\section{Geodesic flows on the superconformal group}

An inner-product $\langle .,.\rangle$ on a Lie algebra $\cg$ determines
a right (or a left) invariant metric on the corresponding Lie group $G$. The 
equation of geodesic motion on $G$ with respect to this metric is 
determined as follows \cite{arn}. Define a bilinear operator 
$\;B:\, \cg\times\cg\ra\cg\;$ by
\be
\Bigl\langle \bigl[  V, W \bigr] \,,\,   U\Bigr\rangle\ =\ 
\Bigl\langle  W \,,\, B(U,V) \Bigr\rangle\quad , \qquad \forall\ \  W\in\cg\ .
\la{B}\ee
The geodesic flow equation is then simply
\be
U_t = B(U,U)\ .
\la{geodesics}\ee
In our case, $\cg$ is the NSR superconformal algebra, consisting of
triples $\,\left(u(x),\f(x),a\right)\,$, where $u$ is a bosonic field, 
$\f$ is a fermionic field and $a$ is a constant. The Lie bracket is given by
\be\begin{array}{l}
 \Bigl[ ( u,\f,a)\,,\, ( v,\j,b)\Bigr] 
\la{alg}\\[6pt]
=\  \left( uv_x {-}u_x v{+}\fr12 \f\j\, ,\, 
               u\j_x {-} \fr12 u_x\j {-}\f_x v {+}\fr12\f v_x\, ,\,  
\displaystyle\int dx  \bigl(c_1 u_x v_{xx} {+} c_2 u v_x {+} {c_1} \f_x \j_x  
               {+} \fr{c_2}4 \f\j \bigr) \right)\, ,
\ea\ee
where $c_1,c_2$ are constants.
On this algebra, an $H^1$ inner product is given by
\bea
\Bigl\langle  ( u,\f,a)\,,\, ( v,\j,b)\Bigr\rangle 
&=& \int dx \left( uv  + \nu u_x v_x 
                      + \a \f\p_x^{-1} \j  + \a\mu \f_x \j \right)\  + ab 
\nonumber \\[6pt]
&=& \int dx\ \left( u\ \D_0\ v  
                      +  \f\ \D_1\ \j  \right)\  +\ ab\quad ,
\la{norm}\eea
where 
\be
\D_0\ =\  1 - \nu \p_x^2 \quad  ,\quad
\D_1\ =\  \a\left(\p_x^{-1} - \mu \p_x\right)\ ,
\ee
and $\,\mu,\nu,\a\,$ are further constants, all assumed nonzero.
Writing $\,U{=}(u,\f,a)\,,\, V{=}(v,\j,b)\,$, we find 
$\,B(U,V)=(B_0, B_1, 0)\,$, where 
\be\arr
\D_0 B_0(U,V)  &=& - \left( 2 v_x \D_0  u + v \D_0  u_x 
                       + \fr32 \j_x \D_1 \f + \fr12 \j \D_1 \f_x \right)
                               + a  (c_1 v_{xxx} - c_2 v_x)\ ,
\\[6pt]  
\D_1 B_1(U,V) 
    &=& - \left( \fr32 v_x \D_1 \f + v \D_1 \f_x + \fr12 \j \D_0 u \right)
                       + a ({c_1} \j_{xx} - \fr{c_2}4 \j)\ .
\ea\la{Brels}\ee
The geodesic equations are therefore conveniently written in the form
\be\arr
\D_0\ u_t  &=& \D_0\ B_0(U,U) \\[6pt]
\D_0\ \f_t  &=& \D_1\ B_1(U,U)\\[6pt]
a_t &=& 0\ .
\ea\ee
Writing $\,\f=\l \x_x\,$, where $\l$ is a constant satisfying 
$\,\l^2={4\over 3\a}\,$, this yields the system
\be\arr
u_t - \nu u_{xxt} 
&=& \k_1 u_x + \k_2 u_{xxx}  - 3 uu_x +  \nu (uu_{xxx}+2u_xu_{xx}) 
        + 2  \x\x_{xx} +  \fr{2\mu}{3} \x_x\x_{xxx}\ ,
\\[6pt]
\x_t - \mu \x_{xxt} 
&=&  \fr{\k_1}{4\a} \x_x +   \fr{\k_2}{\a}  \x_{xxx} 
- \fr32  u_x \x  - (1+\fr{1}{2\a} ) u \x_x
 + \mu u\x_{xxx} + \fr{3\mu}{2}  u_x \x_{xx} + \fr{\nu}{2\a}  u_{xx}\x_x\ .
\la{geo}\ea\ee
Here $\k_1,\k_2$ are independent parameters determined by
$a,c_1,c_2\,$. This is evidently a 5 parameter class of systems
of type \re{gen}.

Setting $\,\x\,$ to zero in \re{geo} yields the CH result of \cite{mis,sh,ko}.
If instead we choose $\mu,\nu$ to vanish, the $H^1$ norm becomes an
$L^2$ norm; then choosing $\k_1$ to be zero and rescaling $\k_2$ to $1$
we obtain the following 1 parameter fermionic extension of KdV:
\be\arr
u_t  &=& u_{xxx}  - 3 uu_x  + 2  \x\x_{xx}\ ,\\[6pt]
\x_t &=&  \fr1{\a}  \x_{xxx}  - \fr32  u_x \x  - (1+\fr{1}{2\a} ) u \x_x\ .
\la{ferkdv}\ea\ee
Modulo rescalings, the superKdV of Mathieu and Manin-Radul 
is obtained  by taking  $\,\a=1\,$. The kuperKdV system arises 
by taking $\,\a=\fr14\,$, the choice made in \cite{ok}. Other values
of the parameters give systems which are not believed to be integrable
(see however \cite{zhang}).

\section{Hamiltonian flows}
Like  KdV, CH has bihamiltonian structure, and this accounts for its 
integrability. We might hope that for some choices of parameters the
system \re{geo} should also have a bihamiltonian structure. One 
hamiltonian structure follows automatically from the geometric origins
of the system \cite{arn}. Explicitly, introducing new variables,  
$\,m = u - \nu u_{xx}\,$ and $\,\eta = \x - \mu \x_{xx}\,$, \re{geo}
takes the form
\be  \pmatrix{m_t \cr \eta_t } 
   = \cp_2 \pmatrix{ \fr{\d\ch_2}{\d m} \cr \fr{\d\ch_2}{\d\eta} } 
\la{hamo}\ee
where 
\be
\cp_2 =   \pmatrix{ \k_2 \p_x^3 + \k_1 \p_x - \p_x m - m \p_x  & 
                     \fr12\p_x \eta + \eta \p_x \cr
                    -\p_x \eta -\fr12 \eta \p_x &
              \fr3{4\a}(\fr{\k_1}4 + \k_2\p_x^2)- \fr{3m}{8} }\ 
\ee 
and the hamiltonian functional is given succinctly by the $H^1$ inner product
on the algebra,
\be
\ch_2\ =\ \fr12\ \Bigl\langle   U\,,\, U \Bigr\rangle 
       \ =\  \fr12 \int dx \left( u^2 +\nu u_x^2                   
               +\fr43(\x_x\x + \mu \x_{xx}\x_x) \right)\quad .
\la{second}\ee   
This generalises the so-called {\it second Hamiltonian structure} of 
KdV and its fermionic extensions \cite{k,m}. Checking \re{hamo} is 
straightforward: the Euler-Lagrange derivatives 
$\fr{\d\ch_2}{\d m},\fr{\d\ch_2}{\d\eta}$ are defined by
\be
\delta\ch_2 = \int\ dx\ \left(
     \frac{\d\ch_2}{\d m}\ \d m + \frac{\d\ch_2}{\d\eta}\  \d\eta
      \right)\ ,
\ee
from which it follows immediately
that $\,\fr{\d\ch_2}{\d m}=u\,$ and $\,\fr{\d\ch_2}{\d\eta}=\fr43 \x_x\,$.
                    
To investigate the possibility of systems amongst \re{geo} having another 
hamiltonian form, we look at systems of the form 
\be  \pmatrix{m_t \cr \eta_t } 
   = \cp_1 \pmatrix{ \fr{\d\ch_1}{\d m} \cr \fr{\d\ch_1}{\d\eta} } \ ,
\la{hamtwo}\ee
where 
\be
\cp_1 = \pmatrix{ \p_x(1{-}\nu \p_x^2) & 0 \cr 
                   0 &  -\fr{\e_1}2 (1{-}\mu \p_x^2)\cr}\ .
\la{p1}\ee
Here $\e_1$ is a constant
and $\ch_1$ is a functional generalising the KdV {\it first Hamiltonian}, 
\bea
\ch_1 &=& \int dx\ \Bigl(-\fr12 u^3 -\fr{\b_3}{2} uu_x^2 
                   - \fr{\k_2}2 u_x^2 +\fr{\k_1}{2} u^2 
           + \fr{\s_1}{\e_1}\x\x_x +\fr{\s_2}{\e_1} \x\x_{xxx}
\Bigr.\nonumber\\[4pt] &&\qquad\quad  \Bigl.   
     + 2  u\x\x_x  + (\g_2-\g_3) u\x_x\x_{xx} + \g_3 u\x\x_{xxx} 
     \Bigr)\ .
\la{first}\eea
This is the most general functional of this type, up to rescalings of $u$ 
and $\x$. Since $\,\d m{=}(1-\nu\p_x^2)\d u\,$, we have 
$\,(1-\nu\p_x^2)\fr{\d\ch_1}{\d m} 
=\fr{\d\ch_1}{\d u}\,$, and similarly  $\,(1-\mu\p_x^2)\fr{\d\ch_1}{\d \eta} 
=\fr{\d\ch_1}{\d \x}\,$. Thus equations \re{hamtwo} take the simple form
\bea      
u_t - \nu u_{xxt} &=& \p_x\left(\fr{\d \ch_1}{\d u}\right) \nonumber\\
    &=& \k_1 u_x +\k_2 u_{xxx} -3 uu_x 
                   + \b_3(2 u_xu_{xx} +  uu_{xxx})
    + 2 \x\x_{xx} + \g_2 \x_x\x_{xxx}+ \g_3 \x\x_{xxxx} 
\nonumber\\[10pt]
\x_t - {\mu} \x_{xxt} 
      &=& \e_1\left(\fr{\d \ch_1}{\d\x}\right) \nonumber\\
      &=& \s_1 \x_x + \s_2 \x_{xxx} + \e_1( u_x \x + 2 u \x_x)
            + \e_1(2\g_3-\g_2) u\x_{xxx} 
            + \fr32\e_1(2\g_3-\g_2) u_x \x_{xx} \nonumber\\
     &&  + \fr12 \e_1 (4\g_3-\g_2) u_{xx}\x_x
         + \fr12 \e_1 \g_3 u_{xxx} \x  \ .
\la{h1}\eea
This is a 10 parameter class of systems of the form \re{gen}.
Comparing with \re{geo}, we see that the only bihamiltonian systems
occur when $\,\{\mu{=}\nu{=}\b_3{=}\g_2{=}\g_3{=}0\,$, $\e_1{=}-\fr32\,$, 
$\s_1{=}\k_1\,$, $\s_2{=}4\k_2\}\,$, which is equivalent to \re{geo} with  
$\,\{\mu{=}\nu{=}0\,,\,\a{=}\fr14\}\,$, i.e. the kuperKdV system. 
Thus, no new bihamiltonian systems arise.

We note that the systems \re{h1} can be obtained from a Lagrangian.
Introducing a potential $f$ defined by $\,u{=}f_x\,$,
they are Euler-Lagrange equations for the functional
\bea
\cl &=& \int dx\  \Bigl(  
\fr12 (f_x-\nu f_{xxx}) f_t + \fr1{\e_1} (\x-\mu \x_{xx}) \x_t 
+ \fr12 f_x^3 + \fr{\b_3}{2} f_x f_{xx}^2  +  \fr{\k_2}2 f_{xx}^2 
- \fr{\k_1}2 f_x^2 
\Bigr.\nonumber\\[4pt] &&\qquad\quad \Bigl.
-\fr{\s_1}{\e_1} \x\x_x - \fr{\s_2}{\e_1} \x\x_{xxx}
- 2  f_x\x\x_x + (\g_3-\g_2) f_x\x_x\x_{xx} - \g_3 f_x\x\x_{xxx}
\Bigr)\ .
\la{l}\eea

\section{Supersymmetric flows}

Define a fermionic superfield $\ \F(x,\q) = s\x + \q u\ $ and superderivative 
$\ D=\der{\q}+ \q\p_x\ $, where $s$ is a nonzero parameter and $\q$ is an
odd coordinate.
The most general superfield equation having component content of the 
form \re{gen} is the $8$ parameter system,
\bea
\left(1-\nu D^4\right)\F_t 
       &=& \k_1 D^2\F  +\k_2 D^6 \F - \fr2{s^2} \F D^3\F  
          + \left(\fr2{s^2}-3\right) D\F D^2\F  
   + \left(\fr{\g_3}{s^2}{+}\b_3\right) D\F D^6\F 
\nonumber\\[8pt] &&  
   - \fr{\g_3}{s^2} \F D^7\F
  + \left(\b_3{+}\fr{\g_3-\g_2}{s^2}\right)  D^2\F D^5\F +  
 \left(\b_2{-}\b_3{+}\fr{\g_2-\g_3}{s^2}\right)  D^3\F D^4\F \ ,
\la{sgen}\eea
where $\,\{\nu,s,\k_1,\k_2,\b_2,\b_3,\g_2,\g_3\}\,$ are parameters. 
The component equations are,
\bea
u_t - \nu u_{xxt} &=&  \k_1 u_x +\k_2 u_{xxx} -3 uu_x 
                   + \b_2 u_xu_{xx} + \b_3 uu_{xxx}
    + 2 \x\x_{xx} + \g_2 \x_x\x_{xxx}+ \g_3 \x\x_{xxxx}\ ,
\nonumber\\[6pt]
\x_t - \nu \x_{xxt} &=&  \k_1 \x_x + \k_2 \x_{xxx} - \fr2{s^2}u_x \x + 
         \left(\fr2{s^2}-3\right)u \x_x
                    + \left(\fr{\g_3}{s^2}+\b_3\right) u\x_{xxx} 
\nonumber\\[4pt] &&
            + \left(\b_2-\b_3+\fr{\g_2-\g_3}{s^2}\right) u_x\x_{xx} 
            + \left(\fr{\g_3-\g_2}{s^2}+\b_3\right) u_{xx}\x_x
            - \fr{\g_3}{s^2} u_{xxx} \x  \ .
\la{sgen_comp}\eea
These systems are by construction invariant under the supersymmetry transformations
\be
\d u\ =\ \t \x_x\ ,\quad  \d\x\ =\ \frac{\t u}{s^2}\ ,
\la{susy}\ee
where $\t$ is an odd parameter.
The superKdV limit, namely $\,\{\nu,\b_2,\b_3,\g_2,\g_3,\k_1\}\,$ all zero, 
yields, modulo rescalings, the one-parameter family of systems studied by 
Mathieu \cite{m}.

By comparing \re{sgen_comp} and \re{h1} it is straightforward to extract
systems which are both supersymmetric and have 
hamiltonian form \re{hamtwo},\re{p1}.
Taking $\,s^2{=}2\,$ in \re{sgen_comp}, 
$\,\{\nu{=}\mu\,$, $\s_1{=}\k_1\,$, $\s_2{=}\k_2\,$, $\e{=}{-}1\}\,$ in \re{h1}, 
and $\,\{\b_2{=}2\b_3\,$, $\b_3=\g_2{-}\fr52\g_3\}\,$ 
in both, we obtain the systems,
\bea
u_t - \nu u_{xxt} &=&  \k_1 u_x +\k_2 u_{xxx} -3 uu_x 
                   + (\g_2-\fr52\g_3) (2u_xu_{xx} + uu_{xxx})
\nonumber\\[2pt] &&    
           + 2 \x\x_{xx} + \g_2 \x_x\x_{xxx}+ \g_3 \x\x_{xxxx}\ , 
\nonumber\\[8pt]
\x_t - \nu \x_{xxt} &=&  \k_1 \x_x + \k_2 \x_{xxx} - u_x \x -2 u \x_x
                    + \left(\g_2-2\g_3\right) u\x_{xxx} 
\nonumber\\[2pt] &&
            + \fr32\left(\g_2-2\g_3\right) u_x\x_{xx} 
            + \fr12\left(\g_2-4\g_3\right) u_{xx}\x_x
            - \fr12\g_3 u_{xxx} \x  \ .
\la{sh1}\eea
These may be expressed in superfield form \re{sgen} with the above choice of
parameters.
The manifestly supersymmetric hamiltonian form is given by
\be
M_t = \widehat\cp_1\  {\d \widehat\ch_1 \over \d M}\quad , \quad  
M = \F - \nu D^4 \F\ , 
\ee
with
\bea
\widehat\cp_1 &=&  D (1 - \nu D^4)\ ,
\\[4pt]
\widehat\ch_1 &=&  \int dx d\q\ 
         \left( \fr{\k_1}{2} \F D\F  - \fr{\k_2}2 D^2\F D^3\F 
         -\fr12 \F (D\F )^2 
\right. \nonumber\\ && \left. \qquad\qquad 
+\fr14\g_3 \F(D^3\F )^2  + \fr14(\g_2-2\g_3) (D\F)^2 D^4\F \right)\ .
\eea
Since the KdV reduction of \re{sh1} (with $\k_1=\g_2=\g_3=0$) is not believed
to be integrable, we have not explored this class of systems further.

In a similar fashion, we may look for choices of parameter sets for which
the geodesic flow equations of section 2 are also supersymmetric. 
Comparing \re{geo} with \re{sgen_comp}, we see that the choice
$\,\{\mu{=}\nu$, $\alpha{=}1\,,\,\k_1{=}0\}\,$ in the former and 
$\,\{s^2{=}\fr43\,$, $\b_2{=}2\nu\,$, $\b_3{=}\nu\,$, 
$\g_2{=}\frac{2\nu}{3}\,$, $\g_3{=}\k_1{=}0\}\,$ 
in the latter, yields the 
two-parameter system of supersymmetric geodesic flow equations:
\bea
u_t - \nu u_{xxt} 
&=&  \k_2 u_{xxx} - 3 uu_x + 2\x\x_{xx} 
            + \nu (uu_{xxx} + 2 u_xu_{xx}) + \fr{2\nu}{3} \x_x\x_{xxx}\ , 
\nonumber\\[6pt]
\x_t -\nu \x_{xxt} 
&=&  \k_2 \x_{xxx} 
- \fr32 (u \x)_x + \nu ( u\x_{xxx} +\fr32 u_x\x_{xx} + \fr12 u_{xx}\x_{x} )\ .
\la{susy_geo}\eea
We shall call this system, with $\,\k_2{=}0\,$ and 
$\,\nu{\not=}0\,$, the {\em supersymmetric Camassa-Holm equation} (superCH). 
In section 5, we present some evidence for its integrability.
The system \re{susy_geo} reduces to superKdV, upon setting $\nu$ to zero, 
and to CH, upon setting $\x$ to zero and translating $u$.

Not surprisingly, the systems \re{susy_geo} arise as geodesic flow equations
precisely when the metric \re{norm} on the NSR superconformal algebra is 
supersymmetric. Then, the calculations of section 2 can be performed
using superfields. Specifically, writing $\,\cu=u+\q\phi\,$ and 
$\,\cv=v+\q\psi\,$, the bracket \re{alg} takes the form
\be
\Bigl[ (\cu,a)\,,\, (\cv,b)\Bigr] 
 =    \left( \cu D^2 \cv-\cv D^2 \cu + \fr12 D\cu D\cv \; , \;
      c_1 \int dxd\q\, D^2 \cu D^3 \cv  \right)
\la{salg}\ee 
and the inner product \re{norm} may be written
\be
\Bigl\langle  (\cu ,a)\,,\, ( \cv,b)\Bigr\rangle =
 \int dxd\q\ \left( \cu D^{-1} \cv + \nu D^2 \cu D\cv \right)\  + ab\ .
\la{snorm}\ee
The superspace bilinear operator $\widehat B$ is given by 
$\ \widehat B\Bigl((\cu,a),(\cv,b)\Bigr)= (\widehat B_0,0)\,$, 
where $\,\widehat B_0\,$ satisfies
\be
 (1-\nu D^4) D^{-1} \widehat B_0 \ =\ 
c_1 a D^5 \cv -\fr32 D^2 \cv (1{-}\nu D^4) D^{-1} \cu
       - \fr12 D\cv (1{-}\nu D^4) \cu - \cv (1{-}\nu D^4) D\cu\ .
\ee
Writing $\,c_1a{=}\k_2\,$ and $\cu=D\F$,
the geodesic flow equations
$\ \left(\cu_t,a_t\right)=\widehat B\Bigl((\cu,a),(\cu,a)\Bigr)\ $
yield
\be
(1-\nu D^4) \Phi_t  
\ =\  \k_2 D^6 \F -\fr32( \F D^3\F  + D\F D^2\F)  
          + \nu\left( D\F D^6\F +\fr12 D^2\F D^5\F + \fr32 D^3\F D^4\F 
            \right) .
\la{schsufe}\ee
We thus recover the subsystem of \re{sgen} having component content \re{susy_geo}.
Equation \re{schsufe} has superfield hamiltonian formulation,
\be
M_t = \widehat\cp_2\  {\d \widehat\ch_2 \over \d M}\quad , \quad  
M = \F - \nu D^4 \F\ , 
\ee
with
\bea
\widehat\cp_2 &=& \k_2 D^5 - \fr12 DMD - D^2M - MD^2\ ,
\\[4pt] 
\widehat\ch_2 &=& \fr12\ \Bigl\langle  (D\F ,0)\,,\, ( D\F,0)\Bigr\rangle\ 
=\  \fr12\ \int dx d\q\   \F DM\  .  
\eea

\section{Painlev\'e integrability of superCH systems}

In this section we investigate, in more detail, 
the supersymmetric geodesic flow system \re{susy_geo}
with $\,\nu{=}1\,$ and $\k_2=0$,
\be\arr
m_t  &=& -2 m u_x -  u  m_x + 2\eta \x + \fr23 \eta_x \x_x\ ,
\quad&m\ =\ u -  u_{xx}\ ,
\\[6pt]
\eta_t &=&  - \fr32  \eta u_x - \fr12 m \x_x - u \eta_x\ ,
\quad&\eta\ =\  \x -  \x_{xx}\ . 
\la{sch}\ea\ee
We shall consider the two simplest possible choices for the grassmann 
algebra in which the fields are valued, viz. algebras with one or two
odd generators. Taking the algebra to be finite dimensional is a very 
convenient tool for preliminary investigations of systems with grassmann 
algebra-valued fields. Manton \cite{manton} recently studied 
some simple supersymmetric classical mechanical systems in this way
and he introduced the term `deconstruction' to denote a component expansion
in a grassmann algebra basis. In \cite{ds} we investigate fermionic 
extensions of KdV in a similar fashion.

\subsection{First deconstruction of superCH}
We first consider the superCH system \re{sch} with fields taking values in 
the simplest grassmann algebra with basis $\{1,\t\}$,
where $\t$ is a single fermionic generator. In this case the fermionic fields
may be expressed as $\;\x=\t\x_1$, $\eta=\t\eta_1\;$, where $\x_1$ and $\eta_1$ 
are standard (i.e. commuting, c-number) functions, as are $u$ and $m$ in this
simple case. Since $\,\t^2=0\,$, the fermionic bilinear terms do not 
contribute and we are left with the system
\be\arr
m_t  &=& -2 m u_x -  u  m_x \ ,
&m\ =\ u -  u_{xx} 
\\[6pt]
\eta_{1t} &=&  - \fr32  \eta_1 u_x - \fr12 m \x_{1x} - u \eta_{1x}\ ,
\qquad&\eta_1\ =\  \x_1 -  \x_{1xx}\ .
\la{fdsch}\ea\ee
Further analysis is simplified by changing coordinates
as described in \cite{s}. Writing $\,m{=}p^2\,$, the first equation of
(\ref{fdsch}) takes the form $\,p_t=(-pu)_x\,$, which suggests 
new coordinates $y_0,y_1$ defined via 
\be
dy_0 = p\,dx-pu\,dt  \  ,\quad  dy_1 = dt \  ,
\ee
or dually, via
\be
\der{x} =  p \der{y_0}\  ,\quad \der{t} =  \der{y_1} - pu \der{y_0}\  .
\ee
Implementing this coordinate change and eliminating the functions $u$ and $\x_1\,$, the
remaining equations for $p$ and $q\equiv\eta_1$ are:
\be
p^2 \dtt{p}'' - 
p(\dtt{p}p''+\dtt{p}'p') + \dtt{p}p'^2 - 2p^3p' - \dtt{p} = 0\ ,
\label{peq}
\ee
\bea
\dtt{q}'' -\frac{3p'}{p}\dtt{q}' -\frac{3\dtt{p}}{2p}q'' + 
\left(\frac{4p'^2}{p^2}-\frac{2p''}{p}-\frac1{p^2} \right)\dtt{q} + 
\left(\frac{15p'\dtt{p}}{2p^2}-\frac{3\dtt{p}'}{p}-\frac{p}2 \right)q'&& 
    \nonumber\\[6pt]
+3\left(\frac{\dtt{p}p''+2p'\dtt{p}'}{p^2}-\frac{4\dtt{p}p'^2}{p^3}-p'  
  \right)q &=& 0\ .
\label{qeq}\eea
Here the dot and prime denote differentiations with respect to $y_1$ and
$y_0$ respectively. We note: (a) thanks to supersymmetry \re{susy},
if $p$ is a solution of \re{peq}, then $q{=}p^2$ 
is a solution of \re{qeq}; and (b) under the substitution
$\;q=p^{3/2}r\;$, \re{qeq} takes the substantially simpler form
\be
\dtt{r}'' +
\left(\frac{p'^2}{4p^2}-\frac{p''}{2p}-\frac1{p^2} \right)\dtt{r}  
 -\frac{p}2 r' -\frac{3p'}4 r  = 0\ .
\ee

\smallskip

\noindent
{\em The system \re{peq},\re{qeq} passes the WTC Painlev\'e  test.}  

\smallskip

\noindent Proof: Equation \re{peq} is a rescaled version of the Associated
Camassa-Holm equation of \cite{s}. Consideration of solutions with 
$\;p(y_0,y_1)\sim p_0(y_0,y_1)\phi(y_0,y_1)^n\;$
near $\phi(y_0,y_1)=0$, for some $n\not=0$, yields $n=-2$ or $n=1$ 
as the possible leading orders of Laurent series solutions.
We need to perform the WTC Painlev\'e test \cite{WTC} for both these
types of series. The first type, namely, Laurent series solutions 
exhibiting double poles on the singular manifold $\phi(y_0,y_1)=0$, 
have already been considered in \cite{hone}.
These take the form
\be
p= \frac{2\phip\dttt\phi}{\phi^2} - \frac{\dttt\phi'}{\phi} + p_2
    + p_3\phi + p_4\phi^2 +\ldots\ ,
\la{poleseries}\ee
where $\phi,p_2,p_4$ are arbitrary functions of $y_0,y_1$, and
\bea
p_3 &=\ {-1\over 2\phip^2\dttt{\phi}^2} 
&\left(\phip^2\dttt{\phi}\dtt{p}_2+\phip\dttt{\phi}^2p_2'-
\left(\phip^2\ddt{\phi}-2\phip\dttt{\phi}\dttt{\phi}'+\phip'\dttt{\phi}^2
\right)p_2
\right. \nonumber\\[2pt]
&& \quad - \left. \left(\phip\dttt{\phi}\ddt{\phi}''
          - \phip\ddt{\phi}\dttt{\phi}''
          - \dttt{\phi}\phip'\ddt{\phi}'
          + \ddt{\phi}\phip'\dttt{\phi}'\right)    \right)\ .
\eea
We have, at present, no explanation of the remarkable symmetry 
of these expressions under interchange of the independent variables.
The second type of solutions have a simple zero on the singular manifold 
$\phi(y_0,y_1)=0$. They take the form
\be
p = \pm \frac{\phi}{\phip} + p_2\phi^2 + p_3\phi^3 + \ldots\ , 
\la{zeroseries}\ee
where $\phi,p_2,p_3$ are arbitrary functions. The verification of
the consistency of both these types of expansions is straightforward. 
This completes the WTC test for equation \re{peq}.

It remains to look at the equation \re{qeq}. Although linear in $q$,
it is {\em not} automatically Painlev\'e. The movable poles and zeros
in $p$ give rise to movable poles in the 
coefficient functions of the linear equation for $q$, and we need
to examine the resulting singularities of $q$.
If $p$ has a pole on $\;\phi{=}0\,$, then near $\;\phi{=}0\;$ we have 
$\;p\sim 2\dttt{\phi}\phip/\phi^2\,$, and equation \re{qeq} takes the form
$$
\dtt{q}''  
+ \left(\frac{6\phip}{\phi}{+}\ldots\right)  \dtt{q}' 
+ \left(\frac{3\dttt{\phi}}{\phi}{+}\ldots\right)   q''  
+ \left(\frac{4\phip^2}{\phi^2}{+}\ldots\right)  \dtt{q} 
+ \left(\frac{11\phip\dttt{\phi}}{\phi^2}{+}\ldots\right)   q'
+ \left(O\left(\frac1{\phi^2}\right)\right)   q 
= 0\ .
$$
Thus the equation has a solution with $\;q\sim \phi^n\;$ if
$\;n(n{-}1)(n{-}2){+}9n(n{-}1){+}15n=0\,$, giving $\;n{=}-4,-2,0\,$. 
It follows that in the case when $p$ is given by the series \re{poleseries},  
no inconsistencies will arise near the double poles of $p$ 
if \re{qeq} has a series solution of the form
\be
q=\frac{q_0}{\phi^4}+\frac{q_1}{\phi^3}+\frac{q_2}{\phi^2}
   +\frac{q_3}{\phi}+q_4+\ldots
\ee
with $q_0,q_2,q_4$ arbitrary. The consistency of such a solution 
can easilly be verified using a symbolic manipulator. Using {\small MAPLE} 
we find that
\be
q_1=\frac{2\phip' q_0-\phip q_0'}{\phip^2}\ .
\ee
The explicit expression for $q_3$ is too lengthy to be given here.

Suppose now that $p$ has a zero on $\phi{=}0$. Near this, 
$\;p\sim\pm\phi/\phip\;$ and equation \re{qeq} has the structure
$$
\dtt{q}''  
- \left(\frac{3\phip}{\phi}+\ldots\right)  \dtt{q}' 
- \left(\frac{3\dttt{\phi}}{2\phi}+\ldots\right)  q'' 
+ \left(\frac{3\phip^2}{\phi^2}+\ldots\right)  \dtt{q} 
+ \left(\frac{15\phip\dttt{\phi}}{2\phi^2}+\ldots\right)  q'
- \left(\frac{12\phip^2\dttt{\phi}}{\phi^3}+\ldots\right)  q
= 0\ .
$$
Thus  \re{qeq} has a solution with $\;q\sim \phi^n\;$ if
$\;n(n{-}1)(n{-}2){-}\fr92n(n{-}1){+}\fr{21}2n{-}12=0\,$, 
giving $n{=}\fr32,2,4$.
The appearance of a half-integer here is not considered a
violation of the Painlev\'e test (see e.g. \cite{frac}). The half integer
value of $n$ gives rise to a series solution of \re{qeq}, near a zero
of $p$, of the form 
\be
q={q_0}{\phi^{\fr32}}+{q_1}{\phi^{\fr52}}+q_2\phi^{\fr72}+\ldots
\la{3/2}\ee
with $q_0$ arbitrary, and $q_1,q_2,\ldots$ determined by $q_0$
(and the arbitrary functions arising in the series \re{zeroseries} 
for $p$). The two integer values of $n$ tell us that we need to
check the consistency of solutions of \re{qeq} taking the form
\be
q=Q_0\phi^2+Q_1\phi^3+Q_2\phi^4+\ldots
\la{24}\ee
with two arbitrary functions $Q_0$ and $Q_2$. This is indeed consistent;
using {\small MAPLE} we obtain
\be 
Q_1\ =\ \pm 2\phip Q_0p_2 - \frac{1}{3\phip^2\dttt{\phi}}\left(
2\phip^2\dtt{Q_0}+2\phip'\dttt{\phi}Q_0 + \phip\dttt{\phi}Q_0'
   + 4\phip\dttt{\phi}'Q_0\right)\ ,
\ee
with the choice of $\pm$ depending on the choice in \re{zeroseries}.
The general solution of \re{qeq} near a zero of $p$, with three 
arbitrary functions, is a linear combination of the series \re{3/2}
and \re{24}. Thus the system \re{peq},\re{qeq} passes the WTC  test.
\hfill$\square$

\smallskip
\noindent
The WTC test is strong evidence for the complete integrability
of the system \re{peq},\re{qeq}. This in turn demonstrates that
superCH indeed has some integrable content.

\subsection{Second deconstruction of superCH}  

We now consider the system \re{sch} with fields taking values in 
a grassmann algebra with two anticommuting fermionic generators,
$\,\t_1\,,\,\t_2\,$. Expanding in the basis $\;\{ 1,\t_1,\t_2,\t_1\t_2\}$, 
\be\arr
&u\ =\ u_0\ +\ \t_1\t_2\, u_{1}\ ,& \x\ =\  \t_1 \x_1\ +\ \t_2 \x_2\ ,
\\[3pt]
&m\ =\ m_0\ +\ \t_1\t_2\, m_{1}\ ,& \h\ =\  \t_1 \h_1\ +\ \t_2\h_2\ ,
\ea\ee
where  the functions $u_0,u_1,m_0,m_1,\x_1,\x_2,\h_1,\h_2$ are
all standard, we obtain the system:
\bea
m_{0t}&=&-2m_0u_{0x}-u_0m_{0x}\ , 
\qquad\qquad\qquad\qquad\qquad~
m_0=u_0-u_{0xx}\ ,
\la{52}\\[4pt]
\h_{it}&=&-\fr32u_{0x}\h_i-\fr12m_0\x_{ix}-u_0\h_{ix}\ ,
\qquad\qquad\qquad~~~
\h_i=\x_i-\x_{ixx}\ , 
\qquad~
i=1,2\ ,
\la{53}\\[4pt]
m_{1t}&=&-2m_1u_{0x}-2m_0u_{1x}-u_0m_{1x}-u_0m_{1x}
\nonumber\\&&
+2(\h_1\x_2-\h_2\x_1) +\fr23(\h_{1x}\x_{2x}-\h_{2x}\x_{1x})\ , 
\qquad
m_1=u_1-u_{1xx}\ .
\la{54}\eea
Supersymmetry \re{susy} tells us that given a solution  $u_0,m_0$ of \re{52}, 
we can solve the remaining equations by taking 
$\;\x_i=\alpha_i u_0\;$, $\;\h_i=\alpha_i m_0\;$ ($i{=}1,2$),
$\;u_1=\beta u_{0x}\;$ and $\;m_1=\beta m_{0x}\,$, where $\alpha_1,\alpha_2,
\beta$ are arbitrary constants.

We handle the system \re{52}-\re{54} 
following the procedure of the previous section. Writing $m_0{=}p^2\,$
and changing coordinates to $y_0,y_1$, the system can be written:
\bea
u_0'\ =\  \left(\frac1{p}\right)^\dt{}\quad, 
\qquad
&&u_0\ =\ p^2-p\left(\frac{\dtt{p}}{p}\right)'\ ,
\la{55}\\[6pt] 
\x_i'\ =\  \frac{3\h_i\dtt{p}}{p^4}-\frac{2\dtt{\h}_i}{p^3}\quad,
&&\x_i\ =\ \h_i + p\left(\frac{3\h_i\dtt{p}}{p^3}-\frac{2\dtt{\h}_i}{p^2}
              \right)' \ ,\qquad i=1,2\ ,
\la{56}\\[6pt]
\left(\frac{m_1}{p^2}\right)^\dt{}\ =\  -(2u_1p)' &+&
\left(\frac{8(\dtt{\h}_1\h_2-\dtt{\h}_2\h_1)}{3p^3} \right)'
 + \left(\frac{4({\h}_1'\h_2-{\h}_2'\h_1)}{3p^3} \right)^\dt{}\ ,
\nonumber\\[6pt] 
&&m_1=u_1-p(pu_1')'\ .
\la{57}
\eea
Applying the WTC  Painlev\'e test to this is a mammoth task,
so instead we consider the Galilean-invariant reduction and apply the
Painlev\'e test at this level. The Galilean-invariant reduction is obtained,
as usual, by restricting all functions to depend on the single variable 
$\,z{=}y_0{-}vy_1\,$ alone. Evidently the first
equations of both \re{55} and \re{57} can be integrated once 
immediately. Then eliminating $u_0$ from \re{55}, $\x_i$ from
\re{56} and $m_1$ from \re{57}, we obtain,
\bea
\left({p'\over p}\right)'  &=&
-{p\over v} + {c_1\over p} -{1\over p^2}\ , 
\la{g0}\\[6pt]
\h_i'''
- \frac{9p'}{2p}\h_i'' &{+}&\!
\left(\frac{11p}{2v}{-}\frac{5c_1}{p}{+}\frac{4}{p^2}{+}\frac{13p'^2}{2p^2}
\right)\h_i' 
{-}\frac{3p'}{p}\left(
\frac{2p}{v}{-}\frac{3c_1}{p}{+}\frac{3}{p^2}{+}\frac{p'^2}{p^2}
\right)\h_i = 0 ,\  i{=}1,2,
\la{g1}\\[6pt]
 u_1'' &{+}& \frac{p'}{p}u_1'+\left(\frac{2p}{v}-\frac1{p^2}\right)u_1
\ =\   d_1 + \frac4{p^3}(\h_1\h_2'-\h_2\h_1')\ ,
\la{g2}\eea
where $c_1, d_1$ are integration constants. 
The equation for $p(z)$  may be integrated again after multiplying
both sides by $p'/p$; this gives
\be
p'^2 = 1 - 2c_1 p +c_2 p^2 - \fr2v p^3\ ,
\la{w0}\ee
where $c_2$ is another integration constant. This equation is well known 
in KdV theory.  Its general solution can be written in terms of the 
Weierstrass $\wp$-function,
\be
p(z) = - 2v \wp(z) + \fr16 c_2 v  \ ,
\ee
where the periods of $\wp$ are determined by the coefficients $c_1,c_2,v$.
Using \re{w0}, the coefficients in \re{g1} can be simplified. 
Further, we know from supersymmetry that this equation has
a solution $\,\eta_i=p^2$. Substituting $\;\eta_i=p^2q_i\;$ the equation 
becomes a second order equation for $q_i'\,$:
\be
q_i'''+\frac{3p'}{2p}q_i''+\left(-\frac{3p}{2v}-\frac{3}{2p^2}+\frac{c_2}2
\right)q'_i\ =\ 0\ ,\qquad i=1,2 \ .
\la{qeq2}\ee
Supersymmetry \re{susy} allows a reduction of the order of \re{g2} as well.
It implies that $\;u_1=p'/p\;$, $\eta_i{=}p^2\,$ is a solution.
So, writing $\;u_1{=}rp'/p\;,\;\eta_i{=}p^2q_i\,$ in \re{g2} yields a 
first order equation for $r'\,$:
\be
r''+\left(c_2p-\frac{4p^2}{v}-\frac1{p}\right)\frac{r'}{p'}\ =\ 
  \frac{p}{p'}\Bigr(d_1 + 4p(q_1q_2'-q_2q_1')\Bigr)\ .
\ee
Multiplying by the integrating factor $p'^2/p$ and integrating, we obtain
\be
r' = \frac{p}{p'^2}\left(d_1p+d_2+4\int(q_1q_2'-q_2q_1')pp'\ dz
\right) \ ,
\la{req}\ee
where $d_2$ is a further constant of integration. 

Thus the Galilean-invariant reduction of the second deconstruction 
of superCH takes the form of the three equations \re{w0},\re{qeq2},\re{req},
to which we now apply the Painlev\'e test. All substitutions hitherto have 
been ones which do not interfere with the test. Equation \re{w0} has 
movable double poles and movable simple zeros. Near a double pole at $z_0$, 
the series solution contains only even powers of $\,(z-z_0)$,
\be
p(z)\ =\ -\frac{2v}{(z-z_0)^2}+\frac{c_2v}{6}
+\frac{12c_1-c_2^2v}{120}(z-z_0)^2
+\frac{\fr{54}{v}+c_2^3v-18c_1c_2}{3024}(z-z_0)^4
+\ldots
\ee
and near a simple zero at $z_0\,$,
\be
p(z)\ =\  \pm (z-z_0) - \fr12 c_1 (z-z_0)^2  \pm\,\fr16 c_2 (z-z_0)^3
   -\fr1{24}(\fr6{v}+c_1c_2)(z-z_0)^4+\ldots\ .
\ee
At both the zeros and poles of $p$, equation \re{qeq2}, which is
just a linear third order ODE, has regular singular points. Checking
Painlev\'e property for this reduces to doing the necessary 
Frobenius-Fuchs analysis at these regular singular points to check that
no logarithmic singularities in the solutions $q_i$ arise. 
Finally, equation \re{req} gives an explicit formula for $r$ involving
two quadratures. Here the necessary analysis involves wrtiting series
expansions for the integrands near the zeros and poles of $p$,
and checking for the absence of $\,1/(z-z_0)\,$ terms, which would give 
rise to logarithms on integration. We do not present all these 
calculations in detail; with the aid of a symbolic manipulator
they are quite straightforward. We conclude that the 
Galilean-invariant reduction of the second deconstruction of superCH
has the Painlev\'e property.

We note, in conclusion, that two of the equations we have encountered 
are interesting variants of the Lam\'e equation:
In \re{qeq2}, the substitution $\;q_i'=p^{-3/4} h_i\;$ yields 
\be
h_i'{}' 
+\fr38\left({p\over v}-{c_2\over 6}+{c_1\over p} -{7\over 2p^2}
\right)h_i\ =\ 0\ ,
\la{l1}\ee
and similarly, on writing $\;u_1=p^{-1/2} k\;$, the homogeneous part 
of \re{g2} takes the form,
\be
k'{}' + \left({3p\over v}-{c_2\over 4}-{3\over 4p^2} \right) k\ =\  0\ . 
\la{l2}\ee
By the arguments above, the latter is integrable by quadratures.

\section{Outlook}

In this paper we have examined fermionic extensions of the Camassa-Holm 
equation. In particular we have identified the superCH system \re{sch}, 
which, for low dimensional grassmann algebras displays some integrability 
properties. Further investigation is needed in order to determine whether 
the superCH system is integrable irrespective of the choice of grassmann 
algebra, and especially for the field theoretically interesting case with 
infinitely many odd generators. We have not been able to find a Lax pair for
superCH. We also note that the peakon solutions 
of the Camassa-Holm equation do not admit supersymmetrisation
(except when the grassmann algebra has just one fermionic generator);
the peakon solutions are weak solutions, with a discontinuity in the 
first derivative, and the action of the supercharge on such functions
gives objects with insufficient regularity properties to be
considered as weak solutions.

Despite these open questions, our work provides a further instance
of integrability arising in the setting of geodesic flow on
a group manifold. It remains a pressing open problem to understand
integrability from this geometric viewpoint. In this context, we should 
mention that the KP (and super KP) systems have yet to be presented as 
geodesic flow equations. If such a presentation exists, it would have
a bearing on the question of whether there is a KP-type higher dimensional 
generalisation of Camassa-Holm (arising in a way similar to that in which 
KP generalises KdV).

\vskip .5cm
\noindent
{\bf Acknowledgements} 

\noindent
We thank the MPI f\"ur Mathematik in den Naturwissenschaften for hospitality 
to J.S. during a visit to Leipzig and the Department of Mathematics and 
Computer Science of Bar-Ilan University for hospitality to C.D. during a visit
to Israel.

\end{document}